\newif\ifsingle

\newif\ifFullVersion

\ifsingle
\documentclass[11pt,draftclsnofoot, onecolumn]{IEEEtran}		
\else		
\documentclass[9pt,final, twocolumn]{IEEEtran}
\fi

\usepackage{times}
\usepackage{amsmath,dsfont}
\usepackage{amssymb,amsthm}
\usepackage{epsfig,verbatim}
\usepackage{subfigure}
\usepackage{setspace}
\usepackage{color}
\usepackage{cite}
\usepackage{epstopdf}
\usepackage{graphics}
\usepackage{accents}
\usepackage{acronym}
\usepackage[bookmarks,colorlinks]{hyperref}
\usepackage{booktabs}
\usepackage{mathtools}
\usepackage{algorithm}
\usepackage{algorithmic}
\usepackage{enumitem}

\newcommand{\myVec}[1]{{\boldsymbol{#1}}}
\newcommand{\myMat}[1]{{\boldsymbol{#1}}}
\newcommand{\mySet}[1]{\mathcal{#1}}

\newcommand{\dF}{d_\mathrm{F}}

\newtheorem{theorem}{Theorem}

\newtheorem{proposition}{Proposition}

\ifsingle

\else

\fi 

\acrodef{adc}[ADC]{analog-to-digital convertor}
\acrodef{cs}[CS]{compressed sensing}
\acrodef{csi}[CSI]{channel state information}
\acrodef{dma}[DMA]{dynamic metasurface antenna}
\acrodef{dtft}[DTFT]{discrete-time Fourier transform}
\acrodef{dnn}[DNN]{deep neural network} 
\acrodef{map}[MAP]{maximum a-posteriori probability}
\acrodef{snr}[SNR]{signal-to-noise ratio}
\acrodef{sinr}[SINR]{signal-to-interference-and-noise ratio}
\acrodef{bs}[BS]{base station} 
\acrodef{em}[EM]{electromagnetic} 
\acrodef{iot}[IOT]{Interent of Things}
\acrodef{mimo}[MIMO]{multiple-input multiple-output}
\acrodef{mse}[MSE]{mean-squared error}
\acrodef{pdf}[PDF]{probability density function}
\acrodef{rv}[RV]{random variable}
\acrodef{fec}[FEC]{forward error correction}
\acrodef{lti}[LTI]{linear time-invariant}
\acrodef{wss}[WSS]{wide-sense stationary}
\acrodef{psd}[PSD]{power spectral density}
\acrodef{ris}[RIS]{reconfigurable intelligent surface}
\acrodef{ser}[SER]{symbol error rate} 
\acrodef{ber}[BER]{bit error rate} 
\acrodef{sgd}[SGD]{stochastic gradient descent} 
\acrodef{isi}[ISI]{intersymbol interference}  
\acrodef{awgn}[AWGN]{additive white Gaussian noise} 
\acrodef{ut}[UT]{user terminal} 
\acrodef{mmw}[mmWave]{millimeter wave}
\acrodef{wpt}[WPT]{wireless power transfer}

\IEEEoverridecommandlockouts

\newcommand{\p}{\mathbf{p}}

\newcommand{\pil}{\mathbf{p}_{i,l}}

\newcommand{\Ail}{{A}_{i,l}}

\title{Near-Field Wireless Power Transfer with
Dynamic Metasurface Antennas
}
\author{  
	\IEEEauthorblockN{Haiyang Zhang, Nir Shlezinger, Francesco Guidi, Davide Dardari, Mohammadreza F. Imani, and Yonina C. Eldar\\
	} 
	\thanks{
		This project has received funding from 
		the European Union’s H2020 research and innovation program under grant No. 646804-ERC-COG-BNYQ, from the Air Force Office of Scientific Research under grant No. FA9550-18-1-0208, and from the Israel Science Foundation under grant No. 0100101.
		H. Zhang and Y. C. Eldar are with the Faculty of Math and CS, Weizmann Institute of Science, Rehovot, Israel (e-mail: \{haiyang.zhang; yonina.eldar\}@weizmann.ac.il). 
		N. Shlezinger is with the School of ECE, Ben-Gurion University, Beer-Sheva, Israel (e-mail: nirshl@bgu.ac.il). 
		F. Guidi is with the National Research Council of Italy, Institute of Electronics, Computer and Telecommunication
Engineering, Bologna, Italy (e-mail: francesco.guidi@ieiit.cnr.it). D. Dardari is with the Department of Electrical, Electronic, and Information Engineering “Guglielmo Marconi” - DEI-CNIT,
University of Bologna, 
Cesena, Italy (e-mail:davide.dardari@unibo.it). 
M. F. Imani is with the School of ECEE, Arizona State University, Tempe, AZ (email: mohammadreza.imani@asu.edu).}

	\vspace{-1.0cm}
	
}
\vspace{-0.75cm}

\begin{document}
	
	\maketitle
	\pagestyle{empty}
	\thispagestyle{empty}
\begin{abstract}

Radio frequency \ac{wpt} enables charging low-power mobile devices without relying on wired infrastructure. Current existing \ac{wpt} systems are typically designed assuming far-field propagation, where the radiated energy is steered in given angles, resulting in limited efficiency and possible radiation in undesired locations. 
When large arrays at high frequencies, such as \ac{dma}, are employed, \ac{wpt} might take place in the radiating near-field (Fresnel) region where spherical wave propagation holds, rather than plane wave propagation as in the far-field.
In this paper, we study \ac{wpt} systems charging multiple devices in the Fresnel region, where the energy transmitter is equipped with an emerging \ac{dma}, exploring how the antenna configuration can exploit the spherical wavefront to generate focused energy beams. In particular, after presenting a mathematical model for \ac{dma}-based radiating near-field \ac{wpt} systems, we characterize the weighted sum-harvested energy maximization problem of the considered system, and we propose  an efficient solution to jointly design the \ac{dma} weights and digital precoding vector. 
Simulation results show that our design generates focused energy beams that are capable of improving energy transfer efficiency in the radiating near-field with minimal energy pollution. 
\end{abstract}

{\textbf{\textit{Index terms---} Radiating near-field, wireless power transfer, beam focusing, dynamic metasurface antennas.}}

\acresetall

\section{Introduction}

Internet of Everything (IoE) is one of the major applications of future 6G wireless communication networks \cite{matthaiou2021road}. The fact that many IoE devices connected to the network are either battery-powered or battery-less \cite{hu2020energy} gives  rise  to the  need  to  energize  them  in a  simple  and  efficient  manner. Radio frequency (RF)  \ac{wpt} is regarded as a promising technology for charging IoE devices, by utilizing RF signals to wirelessly and simultaneously power multiple devices. Compared with the near-field reactive-based \ac{wpt} techniques, such as inductive coupling and magnetic resonance coupling which require the charged device to be very close to the energy source, RF-based \ac{wpt} is capable of charging devices in a more flexible way over longer distances. Hence, RF-based \ac{wpt} presents many potential applications for supporting and prolonging the operation of IoE devices in in-home setups as well as in industrial and commercial settings \cite{CosMas:17}.  

To date, RF \ac{wpt} is mainly studied for charging devices residing in the far-field \cite{zeng2017communications}. In such cases, given the antennas' size, the operational distance between the energizing transmitter and the receivers is larger than the Fraunhofer distance, and thus the radiating wavefront obeys the conventional plane wave model. In such conditions, the transmitter can only direct its energy towards a given angle via beamsteering techniques, resulting in low efficiency and notable energy pollution, i.e., energy radiated at undesired locations. Nonetheless, future wireless 6G systems are expected to support an ecosystem with IoE devices at mmWave bands \cite{saad2019vision} using massive antenna arrays, such as those realized using \acp{dma}, made of configurable radiating metamaterial elements \cite{Yoo2018TCOM, shlezinger2019dynamic, Huang2020holographic, shlezinger2020dynamic,Liaskos_Visionary_2018}. In this case, devices located in distances ranging from a few centimeters to several tens of meters reside in the {\em radiating near-field} region \cite{guidi2019radio,guerra2021near}. Unlike the far-field case, where the EM field is a plane wave, in the radiating near-field region, the EM field is a spherical wavefront.  In such settings, transmitters can generate focused beams \cite{nepa2017near}, which were shown to mitigate interference in multi-user communications \cite{zhang2021beam}, and it was recently envisioned that this capability can facilitate efficient \ac{wpt} with minimal energy pollution \cite{zhang2021near}. This motivates the exploration of the ability to achieve energy focusing using emerging antenna architectures, such as \acp{dma}.   

%


In this work we study radiating near-field \ac{wpt} when the energy transmitter uses a \ac{dma}, quantifying its capability to charge multiple remote devices with minimal energy pollution by forming focused energy beams.  
We first formulate a mathematical model for DMA-based near-field multi-user \ac{wpt} systems, incorporating both the feasible processing of DMAs as well as the propagation of the transmitted EM waves in near-field wireless channels. 
Then, we jointly optimize the digital precoding vector and the DMA weights for maximizing the weighted sum-harvested energy when working in the radiating near-field, while accounting for the specific Lorentzian-form response of metamaterial elements. 
To design the radiating near-field transmission pattern based on the weighted sum-harvested energy maximization objective, we propose an alternating optimization algorithm to deal with the corresponding non-convex optimization problem.
In particular, we provide a closed-form optimal digital precoding solution for a fixed DMA configuration. Then, we recast the  \ac{dma} elements design problem into a Riemannian manifold optimization problem, which we efficiently solve using the Riemannian conjugate gradient approach. 

Simulation results show that our proposed design  concentrates the transmissions to the desired focal points, illustrating its energy focusing capability. We also show that by exploiting the beam focusing capabilities of DMAs, one can intelligently and efficiently charge multiple users according  to  their  priority/requirements with minimal energy pollution.
To the best of our knowledge, this work is the first to study beam focusing for multi-user \ac{wpt}, facilitating simultaneous power charging of multiple energy receivers.


The rest of this paper is organized as follows: Section \ref{sec:Model} models  DMA-based radiating near-field \ac{wpt} systems, and formulates the sum-harvested power maximization problem. Section \ref{sec:Solution} presents an efficient algorithm for tuning the DMA weights, while Section \ref{sec:Sims} provides numerical results. 
Finally, Section \ref{sec:Conclusions} concludes the paper.

We use boldface lower-case and upper-case letters for vectors and matrices, respectively. 
The $\ell_2$ norm, vectorization,  transpose, conjugate, and Hermitian transpose,  are denoted as $\| \cdot \|$,  ${\rm vec}(\cdot)$, $(\cdot)^T$, $(\cdot)^{\dag}$,  and  $(\cdot)^H$,  respectively, and 
$\mathbb{C}$ is the set of complex numbers.


\section{System Model}
\label{sec:Model}

 In this section, we characterize the mathematical model for DMA-based radiating near-field \ac{wpt}. We begin by introducing the DMA transmission model in Subsection \ref{sub:DMA}. Then, we present the near-field wireless channel model in Subsection \ref{sub:model}, and formulate the harvested power maximization problem in Subsection~\ref{sub:problem}.

  	\vspace{-0.1cm}	
 \subsection{Dynamic Metasurface Antennas} \label{sub:DMA}
\ac{dma} is an emerging technology for realizing large scale antenna arrays using reconfigurable metamaterials, whose physical properties such as permittivity and permeability are dynamically adjustable \cite{shlezinger2020dynamic}. These antenna architectures are typically comprised of multiple microstrips, each containing multiple metamaterial elements. The frequency response of each element is independently adjustable by varying its local dielectric properties \cite{Sleasman-2016JAWPL}. For DMA-based transmitters, each microstrip is fed by an RF chain, and the input signal is radiated by all the  elements within the same microstrip \cite{wang2019dynamic}. 
 
 To model the transmission procedure, consider a DMA with $N_d$ microstrips of $N_e$ elements each, i.e., the total number of tunable metamaterial elements is $N\triangleq N_d \cdot N_e$. Letting ${\bf z}_f \in {\mathbb{C}}^{N_d \times 1}$ denote the input signals to the microstrips, the radiated signal, denoted by ${\mathbf{r}}$, can be written as %
\begin{equation} \label{eq: vector_representation}
{\mathbf{r}}= {\mathbf{H Q}}\,  {\mathbf{z}}_f.
\end{equation}
Here, $\mathbf{Q}\in {\mathbb{C}}^{N \times N_d}$ is the configurable DMAs weights, whose entries are
\begin{equation} \label{eq: weighting_matrix}
{\mathbf{Q}}_{(i-1) N_{e}+l, n}=\left\{\begin{array}{ll}
q_{i, l} & i=n \\
0 & i \neq n \, ,
\end{array}\right.
\end{equation}
where $q_{i,l}$ denotes the frequency response of the $l$-th metamaterial element of $i$-th microstrip. These responses satisfy the  Lorentzian form \cite{DSmith-2017PRA, smith2017analysis}, approximated as
\begin{equation}\label{eqn:FreqSel}
q_{i,l} \in \mathcal{Q}\triangleq \left\{\frac{j+e^{j \phi}}{2}| \phi \in [0,2\pi]\right\}, \qquad \forall i,l.
\end{equation}
In addition, $\mathbf{H}$ in \eqref{eq: vector_representation} is a $N \times N$ diagonal matrix with entries
${\mathbf{H}}_{((i-1)N_e+l,(i-1)N_e+l)}=h_{i,l}$,
where $h_{i,l}$ denotes the signal propagation effect of the $l$-th metamaterial element of $i$-th microstrip (inside the microstrip). These coefficients can be written as
%
%
%
$h_{i,l}=e^{-\rho_{i,l}(\alpha_{\rm c}+ j\beta_{\rm c}) }$, 
where $\alpha_{\rm c}$ and $\beta_{\rm c}$ are two constants depending on the characteristic of DMA, and $\rho_{i,l}$ denotes the location of the $l$-th element in the $i$-th microstrip.


   \vspace{-0.1cm}
 \subsection{DMA-based Near-field Channel Model} \label{sub:model}

  \begin{figure}
		\centering	
		\includegraphics[width=0.68\columnwidth]{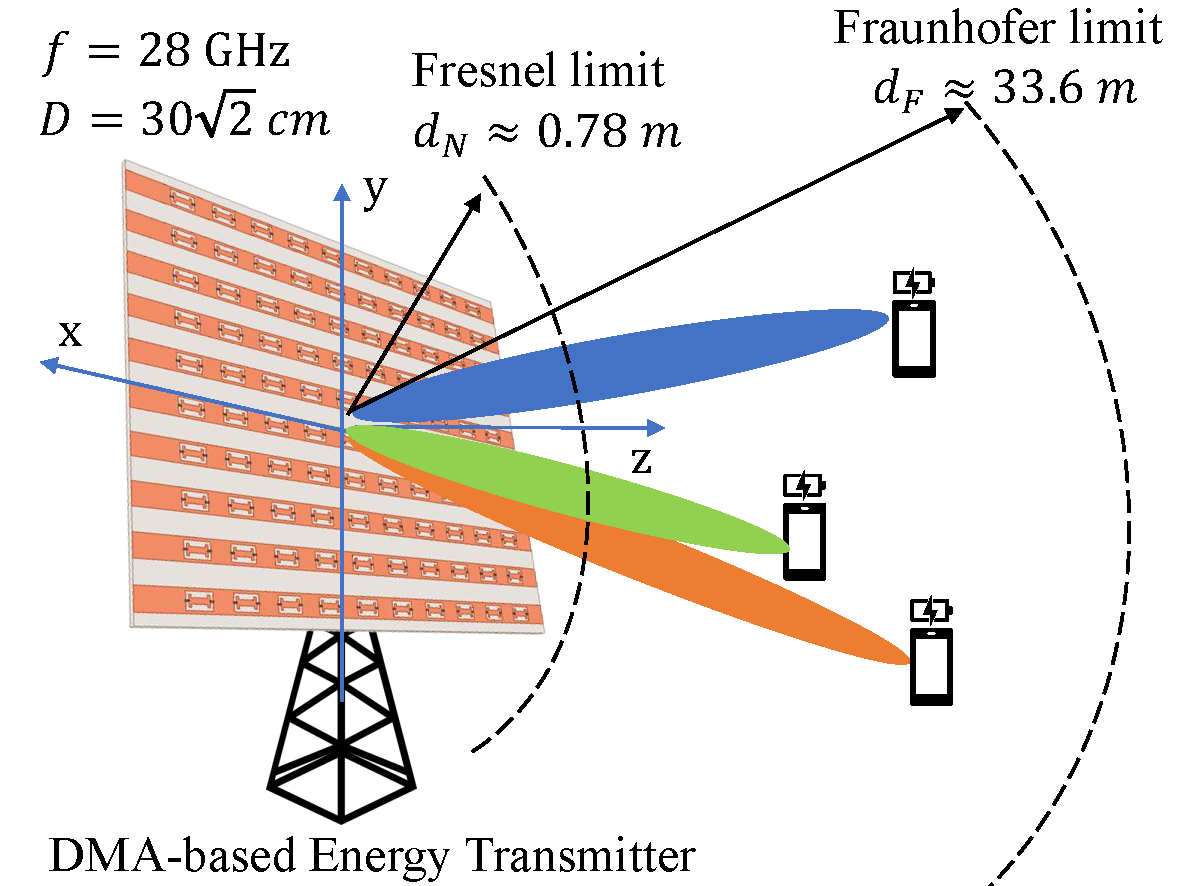}
 		\vspace{-0.2cm}
		\caption{DMA-based energy focusing for radiating near-field multi-user \ac{wpt}.} 
 		\vspace{0.1cm}
		\label{fig:system_model}
	\end{figure}
	
 We consider a radiating near-field multi-user MIMO \ac{wpt} system where a DMA-based energy transmitter charges $M$ single-antenna energy receivers wirelessly, as illustrated in Fig. \ref{fig:system_model}. For the radiating near-field case, the distance between the DMA transmitter and the energy receivers is assumed to be not larger than the Fraunhofer distance $\dF \triangleq \frac{2\,D^2}{\lambda}$ and not smaller than the Fresnel limit $d_{\mathrm N}\triangleq \sqrt[3]{\frac{D^4}{8\,\lambda}}$ \cite{guidi2019radio}, with $D$ and $\lambda$ representing the antenna diameter and the wavelength, respectively. The properties of spherical waves in the radiating near-field allow for the generation of focused beams to facilitate \ac{wpt} \cite{nepa2017near}.

 To formulate the overall energy transmission model, we let $e_m $ be the unit-power energy symbol for the $m$-th energy receiver, $m \in \{1,2,\ldots,M\}\triangleq \mySet{M}$, and use ${\bf w}_m \in {\mathbb{C}}^{N_d \times 1}$ to denote the digital precoding vector. The digital input to the DMA is given by  ${\bf z}_f =\sum_{m=1}^M {\bf w}_m e_m $, and thus by  \eqref{eq: vector_representation} the channel input is 
\begin{equation} \label{eq: vector_representation_1}
\mathbf{r}=\sum_{m=1}^M  \mathbf{H} \mathbf{Q} \, {\bf w}_m e_m \, .
\end{equation}

Let  $\pil=(x_i,y_l,0)$, $i=1,2, \ldots N_d$, $l=1,2, \ldots N_e$, denote the Cartesian coordinate of the  $l$-th element of the $i$-th microstrip. Then, under the free-space condition, the signal received by the $m$-th energy receiver located in $\mathbf{p}_m=(x_m,y_m,z_m)$ can be written as
\begin{equation}\label{eqn:RX1_new}
 	s(\mathbf{p}_m) = \sum_{i=1}^{N_d} \sum_{l=1}^{N_e} \Ail (\p_m)\, e^{ -\jmath k d_{i,l,m}} \,y_{i,l}\, +n_m.
 \end{equation} 
Here, $d_{i,l,m}=|\p_m-\pil|$ is the distance between the  $l$-th element of the $i$-th microstrip and the $m$-th energy receiver; $k \triangleq 2\pi /\lambda$ denotes the wave number;  
$n_m \sim \mathcal{C} \mathcal{N}\left(0, \sigma^{2}\right)$ is white Gaussian noise; and $\Ail(\p_m)$ is the path-loss coefficient. Following \cite{ellingson2019path}, we have 
%
$\Ail(\p_m)=\sqrt{F(\Theta_{i,l,m})}\frac{\lambda}{4\,\pi d_{i,l,m}}$,
where $\Theta_{i,l,m}=(\theta_{i,l,m},\phi_{i,l,m})$  is the elevation-azimuth pair from the $l$-th element of the $i$-th microstrip to the $m$-th energy receiver, and $F(\Theta_{i,l,m})$ is the radiation profile modeled as
\begin{align} \label{eqn:radiationProfile}
    F(\Theta_{i,l,m}) \!=\! \left\{\begin{array}{ll} 2\, (b+1)\, \cos^b (\theta_{i,l,m})  & \,  \theta_{i,l,m} \in [0,\pi/2] \, ,   \\0 & \, \text{otherwise}. \\ \end{array}\right.    
 \end{align}
 In \eqref{eqn:radiationProfile}, the parameter $b$ is the Boresight  gain constant, e.g., $b=2$ for the dipole case \cite{ellingson2019path}.



For ease of analysis, we rewrite \eqref{eqn:RX1_new} in the following compact form $s(\mathbf{p}_m) = {\bf a}_m^H\,\mathbf{r} +n_m$, with  ${\bf a}_m \triangleq \big[A_{1,1}(\p_m)\, e^{ -\jmath k d_{1,1,m}},\ldots , A_{N_d,N_l}(\p_m)\, e^{ -\jmath k d_{N_d,N_l,m}}\big]^H $. Then, 
by using the expression for the channel input $\mathbf{y}$ given in \eqref{eq: vector_representation_1}, the received signal of the $m$-th energy receiver is given by
\begin{equation} \label{eqn:RX2_vector}
s(\mathbf{p}_m) ={\bf a}_m^H\,\sum_{j=1}^M \mathbf{H} \mathbf{Q} \, {\bf w}_j x_j +n_m, \quad \forall m \in \mySet{M}.
\end{equation}

  \vspace{-0.1cm}
 \subsection{Problem Formulation} \label{sub:problem}
  
Using the channel formulation \eqref{eqn:RX2_vector} and the energy harvesting model proposed in \cite{xu2014multiuser}, the harvested power from the transmitted signal of the $m$-th energy receiver s given by 
\begin{equation} \label{eq:total-harvested power}
E_m = \zeta\, \sum_{j =1 }^M\left|{\bf a}_m^H\, \mathbf{H} \mathbf{Q}\, {\bf w}_j \right|^{2}, \quad m \in \mySet{M},
\end{equation}
%
where $0 < \zeta <1$ is the energy conversion efficiency. 

Our aim is to design a transmission scheme, including both the digital precoding as well as the \ac{dma} configuration, to enable multi-user \ac{wpt} in the radiating near-filed region.  This is expressed as the joint optimization of the DMA weights $\mathbf{Q}$ and the digital digital precoding vectors $\left\{{\bf w}_m \right\}$ to maximize the weighted sum-harvested energy, subject to both the total transmit power constraint $P_{\max}$, and the structure constraint on the DMA weights matrix $\mathbf{Q}$ in \eqref{eq: weighting_matrix}.  Mathematically, the problem of interest can be formulated as
\begin{equation} \label{eq:optimization_problem1}
\begin{split}
&\max_{ \left\{{\bf w}_m\right\},{\bf Q}}~~\zeta ~\sum_{m=1}^{M} \alpha_m E_m 
\\
&~~s.t.~~~~~~\eqref{eq: weighting_matrix}, \quad q_{i, l} \in \mathcal{Q}, \forall i,l, \quad \sum_{m=1}^{M} \left\|{\bf HQ w}_m\right\|^2 \leq P_{\rm max},
\end{split}
\end{equation} 
where $\{\alpha_m\}_{m=1}^M$, are predefined weights that are application-specific. 

	\section{DMA Beam Focusing for WPT}
	\label{sec:Solution}

In this section, we study the joint design of the digital precoding vector and the DMA weights for maximizing the weighted sum-harvested energy. Note that 
 \eqref{eq:optimization_problem1} is non-convex due to the coupled optimization variables in both the objective function and constraints, as well as the Lorentzian constraints on metamaterial elements. To make \eqref{eq:optimization_problem1} more tractable, we relax it as follows
\begin{equation} \label{eq:optimization_problem}
\begin{split}
&\max_{ \left\{{\bf w}_m\right\},{\bf Q}}~~\zeta ~\sum_{m=1}^{M}  \sum_{j =1 }^M  \alpha_m\left|{\bf a}_m^H\, \mathbf{H} \mathbf{Q}\, {\bf w}_j \right|^{2}\\
&~~s.t.~~~~~~\eqref{eq: weighting_matrix}, \quad q_{i, l} \in \mathcal{Q}, \forall i,l, \quad \sum_{m=1}^{M} \left\|{\bf w}_m\right\|^2 \leq P_{\rm max}.
\end{split}
\end{equation} 
%
 The problem \eqref{eq:optimization_problem} differs from \eqref{eq:optimization_problem1} in its power constraint, which is imposed on the digital output rather than on the transmitted signal. However, one can derive the digital precoder based on \eqref{eq:optimization_problem}, and scale $\{\myVec{w}_m\}$ such that the transmitted  power constraint in \eqref{eq:optimization_problem1} holds. 
 
Since problem \eqref{eq:optimization_problem} is still non-convex, we propose to individually optimize $\myMat{Q}$ and $\{\myVec{w}_m\}$ in an alternating manner.  In the following, we show how to solve \eqref{eq:optimization_problem} for fixed $\myMat{Q}$ and for fixed $\{\myVec{w}_m\}$, respectively. Due to page limitations, the proofs of the results can be found in   \cite{Longerversion1}.  

	\vspace{-0.1cm}
\subsection{Optimizing the Digital Precoder}
When $\bf Q$ is fixed, \eqref{eq:optimization_problem} reduces to the  weighted sum-harvested energy maximization problem in multi-user \ac{wpt} systems.  By defining ${\bf G}\left(\mathbf{Q}\right)=\zeta ~\sum_{m=1}^{M} \alpha_m\,\mathbf{Q}^H\,\mathbf{H}^H\,{\bf a}_m\,{\bf a}_m^H\, \mathbf{H} \mathbf{Q}$, the weighted sum-harvested energy  can be reformulated as $\sum_{j=1}^{M} {\bf w}_j^H\,{\bf G}\,{\bf w}_j$. As a result, for a fixed $\bf Q$, \eqref{eq:optimization_problem} is transformed into
\begin{equation} \label{eq:sub1}
\max_{ \left\{{\bf w}_j\right\}}~~\sum_{j=1}^{M} {\bf w}_j^H\,{\bf G}\left(\mathbf{Q}\right)\,{\bf w}_j, \quad
~s.t.~~~ \sum_{j=1}^{M} \left\|{\bf w}_j\right\|^2 \leq P_{\rm max}.
\end{equation} 

Following \cite{xu2014multiuser}, we have the following proposition, which provides the closed-form optimal solution to \eqref{eq:sub1}.
\begin{proposition}
\label{prop:digital_solution}
Let  ${\bf w}^*\left(\mathbf{Q}\right)$ be the eigenvector corresponding to the maximal eigenvalue  of ${\bf G}\left(\mathbf{Q}\right)$. Then, \eqref{eq:sub1} is maximized by setting ${\bf w}_j = \sqrt{p_j}{\bf w}^*\left(\mathbf{Q}\right)$ for any non-negative $\{p_j\}$ s.t. $\sum_{j=1}^M p_j = P_{\rm max}$.
\end{proposition}

Proposition \ref{prop:digital_solution} indicates that all digital precoding vectors share the same transmission direction as ${\bf w}^*\left(\mathbf{Q}\right)$, and the total transmit power should be used to maximize the weighted sum-harvested energy. Without loss of generality, we henceforth set the digital precoder for a given $\myMat{Q}$ to be
\begin{equation}  \label{eq:sub1_solution}
{\bf w}_1 = \sqrt{P_{\rm max}} {\bf w}^*\left(\mathbf{Q}\right),~ \text{and}~ {\bf w}_2=\cdots={\bf w}_M = 0.
\end{equation}
From \eqref{eq:sub1_solution} we see that a single digital precoding vector is sufficient to maximize the weighted sum-harvested energy for a given $\myMat{Q}$. This is because energy symbols do not carry information, thus each receiver can harvest energy from the same symbol. 

	\vspace{-0.1cm}
\subsection{Optimizing the DMA Weights} 

We next focus on solving \eqref{eq:optimization_problem} for fixed $\left\{{\bf w}_j\right\}$. According to \eqref{eq:sub1_solution}, problem \eqref{eq:optimization_problem} for fixed $\left\{{\bf w}_j\right\}$ is simplified as
\begin{equation} \label{eq:sub2}
\max_{{\bf Q}}~~\zeta ~\sum_{m=1}^{M} \alpha_m\,\left|{\bf a}_m^H\, \mathbf{H} \mathbf{Q}\, {\bf w}_1 \right|^{2}, \quad s.t.~~\eqref{eq: weighting_matrix}, q_{i, l} \in \mathcal{Q}, \forall i,l.
\end{equation} 
%
%
%
To proceed, we define the $ N_d^2 \cdot N_e  \times 1$ vectors  ${\bf q}={\rm vec}\left(\bf Q \right) $, and ${\bf z}_m=\left({\bf w}_1^T \otimes ({\bf a}_m^H\, \mathbf{H})\right)^H$. Using these definitions, we identify an equivalent optimization problem to problem \eqref{eq:sub2}, as stated in following theorem.
%
\begin{theorem}
\label{thm:MultiUser}
  For fixed ${\bf w}_1$, \eqref{eq:sub2} is equivalent to:
\vspace{-0.1cm}
\begin{equation} \label{eq:simplifiedx} 
\min_{ {\bf \bar q}}~~  {\bf \bar q}^H\, {\bf A}\left({\bf w}_1\right)\, {\bf \bar q}, \quad 
s.t.~~~{\bar q}_{l} \in \mathcal{Q},~\forall l \in  \mathcal{A}_q, 
\end{equation} 
where $\mathcal{A}_q$ is the set of all non-zero elements of ${\bf q}$, ${\bf \bar q}$ is the modified version of ${\bf q}$ obtained by removing all the zero elements of  ${\bf q}$; ${\bf A}\left({\bf w}_1\right) \triangleq - \zeta \sum_{m=1}^{M} \alpha_m {\bf \bar z}_m\, {\bf \bar z}_m^H$, with 
${\bf \bar z}_{m}$ being the modified version of ${\bf z}_m$ obtained by removing the elements having the same index as the zero elements of ${\bf q}$.
\end{theorem}

The equivalence between  \eqref{eq:sub2} and  \eqref{eq:simplifiedx} holds in the sense that  they achieve the same optimal value. Thus, the solution to \eqref{eq:sub2} can be recovered from that of  \eqref{eq:simplifiedx} according to the structure of ${\bf Q}$ \eqref{eq: weighting_matrix}.

Problem \eqref{eq:simplifiedx} is still non-convex and includes the Lorentzian constraint ${\bar q}_{l} \in \mathcal{Q}$ defined in \eqref{eqn:FreqSel}. This constraint characterizes the feasible set as a circle on the complex plane $\left|{\bar q}_{l} - \frac{1}{2}e^{j \frac{\pi}{2}}\right| = \frac{1}{2}$, with the circle center at $(0,\frac{1}{2}e^{j \frac{\pi}{2}})$ and radius equal to $\frac{1}{2}$. In order to simplify  \eqref{eq:simplifiedx}, we define a new vector variable ${\bf b} \in \mathbb{C}^{N}$ whose $l$-th entry is given by 
\begin{equation}  \label{eq:change_variable}
b_l=2{\bar q}_{l} - e^{j \frac{\pi}{2}}, \quad \forall l \in  \mathcal{A}_q.
\end{equation}
The variable $b_l$ lies on the unit circle of complex plane, i.e., $\left|b_l\right| = 1$. According to \eqref{eq:change_variable}, we have ${\bf \bar q}=\frac{1}{2}\left({\bf b}+e^{j \frac{\pi}{2}}{\bf 1}\right)$, where ${\bf 1}$ denotes  a $N \times 1$ all ones vector. Hence, we transform 
 \eqref{eq:simplifiedx} into 
\begin{equation} \label{eq:changed}
\begin{split}
&\min_{ {\bf b}}~~f\left({\bf b}\right) \triangleq \frac{1}{4} \left({\bf b}+e^{j \frac{\pi}{2}}{\bf 1}\right)^H {\bf A}\left({\bf w}_1\right)\, \left({\bf b}+e^{j \frac{\pi}{2}}{\bf 1}\right)\\
&~~s.t.~~\left|b_{l}\right| =1,~\forall l \in  \mathcal{A}_q.
\end{split}
\end{equation} 
The search space in \eqref{eq:changed} is the product of $N$ complex circles, which is a Riemannian submanifold of ${\mathbb{C}}^N$. Thus, \eqref{eq:changed} can be tackled using the Riemannian conjugate gradient (RCG) algorithm \cite{yu2019miso,zhang2021beam}.

Denote by  ${\bf Q}$ and ${\bf w}_1$ as the optimal solution to problem \eqref{eq:optimization_problem}. Then, we can scale ${\bf w}_1$ to ${\bf w}_1=\sqrt{P_{\rm max}}\frac{{\bf w}_1}{\left\|{\bf H Q}{\bf w}_1\right\|}$ such that the resulting new ${\bf w}_1$ together with ${\bf Q}$ are an effective approximate solution to problem \eqref{eq:optimization_problem1}, satisfying the transmitted signal power constraint.

Our proposed alternating approach for solving problem \eqref{eq:optimization_problem1} is summarized as Algorithm \ref{algorithm2}. In particular, 
in the $4$th step, the
updating of ${\bf b}^{\left(t+1\right)}$ through RCG algorithm envolves both ${\bf b}^{\left(t\right)}$ in step 3 as its initial value, and the Euclidean gradient of  the objective $f\left({\bf b}\right)$ at point ${\bf b}$, that is, $ \nabla\,f\left({\bf b}\right) =\frac{1}{2} \left({\bf A}\left({\bf w}_1\right)\,{\bf b} + e^{j \frac{\pi}{2}}\,{\bf A}\left({\bf w}_1\right)\, {\bf 1}\right) $, for the calculation of the Riemannian gradient.

%
%
%
 \begin{algorithm}[t!]
 \caption{Proposed algorithm for solving problem  \eqref{eq:optimization_problem1}}
 \label{algorithm2}
 \begin{algorithmic}[1]
  \renewcommand{\algorithmicrequire}{\textbf{Initialize:}} \REQUIRE  ${\bf Q}^{\left(0\right)}$; \\
  \FOR{$t=0,1,\ldots,T$ }
  \STATE Calculate ${\bf w}_1^{\left(t\right)}$ based on \eqref{eq:sub1_solution}, and then update ${\bf A}\left({\bf w}_1^{\left(t\right)}\right)$; \\
  \STATE Calculate ${\bf b}^{\left(t\right)}$ based on ${\bf Q}^{\left(t\right)}$ and  \eqref{eq:change_variable}; \\
  \STATE Update ${\bf b}^{\left(t+1\right)}$ by solving  problem \eqref{eq:changed};\\
    \STATE Obtain ${\bf \bar q}^*$ for problem \eqref{eq:simplifiedx} based on ${\bf b}^{\left(t+1\right)}$ and \eqref{eq:change_variable};\\
  \STATE Update ${\bf Q}^{\left(t+1\right)}$ for problem \eqref{eq:sub2} based on ${\bf \bar q}^*$ and \eqref{eq: weighting_matrix};\\
  \STATE $t=t+1;$
  \ENDFOR
    \STATE ${\bf w}_1^*=\sqrt{P_{\rm max}}\frac{{\bf w}_1^{\left(T\right)}}{\left\|{\bf H}{\bf Q}^{\left(T\right)}{\bf w}_1^{\left(T\right)}\right\|}$; 
 \renewcommand{\algorithmicrequire}{\textbf{Output:}} \REQUIRE  ${\bf w}_1^*$,  ${\bf Q}^{*}={\bf Q}^{\left(T\right)}$.
 \end{algorithmic} 
 \end{algorithm}
	\vspace{-0.1cm}
 \subsection{Discussion}
  \label{sec:Discussion}

The considered weighted sum-harvested energy is also a commonly used metric in conventional far-field multi-user \ac{wpt} scenarios \cite{zeng2017communications}. 
While we do not explicitly enforce the DMA to generate focused energy beams, this indeed happens when seeking to maximize the weighted sum-harvested energy, as numerically illustrated in Section \ref{sec:Sims}. This is because we here consider the radiating near-field scenario, where the energy beam focusing capability inherently exists, and is implicitly encapsulated in the objective via  $\{{\bf a}_m\}$.
As shown  in Section \ref{sec:Sims}, such 
energy focusing  brings forth several advantages to radiating near-field WPT systems. First, it enables enhancing the energy transfer efficiency compared with directive radiation in the far-field. Second, 
it reduces energy pollution and limits human exposure to radiated energy. Therefore, this capability is expected to notably facilitate the charging of 6G IoE devices in indoor settings.

For multi-user wireless communications operating in radiating near-field region, beam focusing has been exploited  to mitigate co-channel interference and hence maximize the sum-rate in our previous work \cite{zhang2021beam}. Despite the similarity between multi-user  near-field \ac{wpt} (considered here) and communications (considered in \cite{zhang2021beam}), there are several fundamental differences in both the design objectives and proposed algorithms. Specifically, for wireless communications,  focused beams are designed to reduce co-channel interference which is harmful to data transmission rate. In \ac{wpt}, co-channel interference is a useful energy source for energy receivers, resulting in different focused beams design considerations to fit different objectives. The fact that beam focusing designs differ between \ac{wpt}  and wireless communications motivates exploring  simultaneous wireless information and power transfer, which is a paradigm allowing a hybrid information and energy transmitter to communicate and power multiple devices at the same time \cite{xu2014multiuser}, in the  radiating near-field.  
However, we leave this extension for future work.

	\section{Numerical Evaluations}
	\label{sec:Sims}


In this section, we present some representative numerical results to demonstrate the potential of energy beam focusing for radiating near-field \ac{wpt}. We consider a radiating near-field WPT system where the energy transmitter is equipped with a planar DMA positioned in the $xy$-plane, and the single-antenna energy receivers are positioned in the $xz$-plane. The antenna size is $30~ {\rm cm} \times 30~ {\rm cm}$. The inter-element spacing in DMA is  $\lambda / 2$, and the numbers of microstrips and metamaterial elements are $N_d = N_e = \lfloor 2D/\lambda\rfloor$, where $\lfloor \cdot\rfloor$ is the integer floor function. We use   { $\alpha_{\rm c} = 1.2~[{\rm m}^{-1}]$ and $\beta_{\rm c} = 827.67 ~[{\rm m}^{-1}]$} \cite{zhang2021beam}, to represent the propagation inside the  waveguides.  We set $P_{\max}$ to be $1~$W,  and the RF-to-DC energy conversion efficiency  is   $\zeta=0.5$.

\begin{figure} 
  \centering 
  \subfigure[Near-field WPT]{ 
    \label{fig:subfig:near-field}
    \includegraphics[width=2.1in]{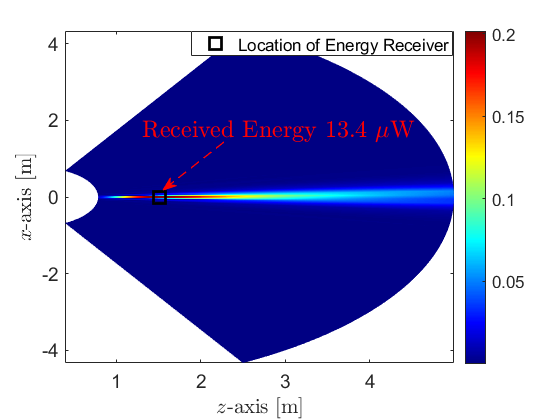} 
  } 
  \subfigure[Far-field WPT]{ 
    \label{fig:subfig:far-field} 
    \includegraphics[width=2.1in]{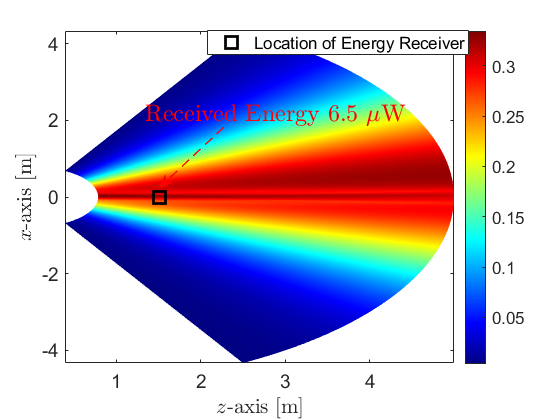} 
  } 
		\vspace{-0.3cm}
  \caption{The normalized received power of the energy receiver located at the: (a) near-field region; (b) far-field region.} 
  \label{fig:rate_twoUser} 
\end{figure}

To demonstrate the gains of near-field energy beam focusing over far-field beam steering, Fig.~\ref{fig:rate_twoUser} depicts the numerically evaluated normalized received power at each point of the predefined region in the $xz$-plane, where the normalized received power is defined as the ratio of the received power of an energy receiver to its corresponding channel gain for removing the influence of path-loss.  The energy transmission scheme is designed  to maximize the received power of the single energy receiver located at ${\rm F}_1(x,y,z)=(0,0,1.51~{\rm m})$. In Fig. \ref{fig:rate_twoUser}(a), we set the frequency as $28~$GHz, so that the location of the target energy receiver is in the radiating near-field region; whereas in Fig. \ref{fig:rate_twoUser}(b), we set the carrier 
frequency as $1.2~$GHz, resulting  in  the  target energy receiver being located in the far-field region. It is observed from Fig. \ref{fig:rate_twoUser}(a) that, in the near-field case, the energy beam is focused around the target energy receiver area, and the power harvested by the target energy receiver is up to $13.4~\mu W$. By contrast, for the far-field case as shown in Fig. \ref{fig:rate_twoUser}(b), energy can only be transmitted towards a direction with a comparatively wider energy beam. Consequently, far-field signalling results in the target energy receiver harvesting only $6.5~\mu W$, which is only $48\%$ of the power obtained in the near-field.  
This gain is achieved despite the fact that the system in Fig. \ref{fig:rate_twoUser}(b) operates at a lower frequency and hence with a lower isotropic path loss.
Besides, by comparing Figs. \ref{fig:rate_twoUser}(a) and \ref{fig:rate_twoUser}(b), it is observed that for the radiating near-field WPT system, energy beam focusing is capable of not only enhancing energy transfer efficiency, but also reducing energy pollution. 

		\begin{table}
		\centering
		\caption{A comparison of harvested energy of each energy receiver under different combinations of weighting coefficients.}\label{tabel1}
		\begin{tabular}{ |c|c|c| } 
			\hline
			Received Energy  & Energy Receiver 1 & Energy Receiver 2\\
			\hline
			$\alpha_1=0.5$, $\alpha_2=0.5 $ & 30.3 $\mu W$  & 2.5 $\mu W$  \\
			\hline
			$\alpha_1=0.1$, $\alpha_2=0.9 $ & 18.7 $\mu W$& 4.7 $\mu W$\\
			\hline
		\end{tabular}
	\end{table}

In Table~\ref{tabel1}, we show the received power of two energy receivers incurred by our proposed Algorithm \ref{algorithm2} under different combinations of weighting coefficients. The energy receivers are located at ${\rm F}_1(x,y,z)=(0,0,0.97~{\rm m})$ and ${\rm F}_2(x,y,z)=(0,0,1.51~{\rm m})$, lying in a similar angular direction. It is observed from Table~\ref{tabel1} that for the case of $\alpha_1=0.5$,  $\alpha_2=0.5 $, the harvested power of energy receiver 1 is much larger than that of energy receiver 2. This is because energy receiver 1 has a better channel condition and thus energy beams are mainly focused on around its location to maximize the objective for the case of having the same weighting coefficient.  When we change the weighting coefficients to $\alpha_1=0.1$, $\alpha_2=0.9 $, the power harvested by  energy receiver 2 increases from   $2.5~\mu $W to  $4.7~\mu $W, while  the power harvested by the energy receiver 1 decreases from  $30.3~\mu $W to  $18.7~\mu $W. This is because the energy transmitter is capable of intelligently charging multiple users according to their priority/requirements even if multiple energy receivers have similar angular direction, thanks to the distinguishing capability of the near-field energy focusing. We point out that beam steering in the far-field does not possess such distinguishing ability, which  is especially important for future 6G IoE applications where  devices  are expected to be densely deployed in the Fresnel region.

	\section{Conclusions}
	\label{sec:Conclusions}
	In this work we studied the use of DMAs for multi-user \ac{wpt} in the radiating near-field region. We presented a model for  DMA-based radiating near-field WPT systems. We then formulated the joint optimization of the DMAs weights and digital precoders  to maximize the weigthed sum-harvested energy, and  proposed  efficient algorithms to solve the resulting non-convex problems. Numerical results demonstrated that using DMAs for energy focusing results in improved energy transfer efficiency in the radiating near-field with minimal energy pollution. 

	\bibliographystyle{IEEEtran}
	\bibliography{IEEEabrv,refs}

\end{document}